\documentclass[12pt]{article}

\usepackage{amsmath,amssymb,array,amsfonts}
\usepackage[dvipdfmx]{graphicx}
\usepackage{comment,color}
\usepackage{cite}

\newcommand{\ba}{\begin{eqnarray}}
\newcommand{\ea}{\end{eqnarray}}

\allowdisplaybreaks

\def\ncm{\newcommand}

\def\s {{\rm s}}

\def\SM{{\rm SM}}

\def\dis{\displaystyle}

\def\nt{\notag}

\ncm{\sls}[1]{{\ooalign{\hfil/\hfil\crcr$#1$}}}

\makeatletter
    
    \@addtoreset{equation}{section}
\makeatother

\begin{document}
\setlength{\baselineskip}{18pt}
\begin{titlepage}

\begin{flushright}
OCU-PHYS 418
\end{flushright}
\vspace{1.0cm}
\begin{center}
{\Large\bf $H \to b \bar{b}, \tau \bar{\tau}$ 
in Gauge-Higgs Unification} 
\end{center}
\vspace{25mm}

\begin{center}
{\large
Yuki Adachi 
and 
Nobuhito Maru$^{*}$
}
\end{center}
\vspace{1cm}
\centerline{{\it
Department of Sciences, Matsue College of Technology,
Matsue 690-8518, Japan.}}

\centerline{{\it
$^{*}$
Department of Mathematics and Physics, Osaka City University, Osaka 558-8585, Japan.
}}
%
%
\vspace{2cm}
\centerline{\large\bf Abstract}
\vspace{0.5cm}
We study the deviation of Higgs production through the gluon fusion process 
 and the decay of Higgs boson to the bottom quark and the tau lepton 
 in gauge-Higgs unification from the Standard Model prediction. 
We find that the signal strength of $gg \to H \to b\bar{b}, \tau\bar{\tau}$ is necessarily suppressed 
 comparing to the SM predictions due to the dominant suppression of the Higgs production via gluon fusion. 
\end{titlepage}

\newpage
\section{Introduction}

A Higgs boson was discovered at the CERN Large Hadron Collider (LHC) experiment \cite{Higgs}, 
 but the couplings of the Higgs boson to the Standard Model (SM) fields 
 and the self-couplings of Higgs boson have not been precisely measured.  
An issue whether the Higgs boson is the SM one or that of physics beyond the SM is still open. 
Physics beyond the SM is expected to exist by several reasons such as the hierarchy problem.

Gauge-Higgs unification (GHU) \cite{GH}, which is one of the attractive scenarios beyond the SM, 
 solves the hierarchy problem without supersymmetry \cite{HIL}. 
In this scenario, 
 the SM Higgs boson is identified with extra spatial components of a higher dimensional gauge field.
A remarkable fact is that the quantum correction to Higgs mass (and potential) is calculable 
 due to the higher dimensional gauge symmetry 
 although the theory is non-renormalizable.
The finiteness of the Higgs mass has been verified in various types of models 
 at one-loop level \cite{ABQ} and at two loop level \cite{MY}. 
The finiteness of other physical observables have been investigated 
 by the present authors or one of them \cite{LM, Maru, ALM}.

The fact that the Higgs boson is a part of gauge field implies that
 Higgs interactions are governed by the gauge principle
 and specific predictions in LHC and ILC physics are expected. 
In fact, the diphoton and $Z\gamma$ decay of the Higgs boson 
 was studied in the framework of GHU and remarkable predictions were found \cite{MO}. 
In order to explain experimental results of diphoton decay and 126 GeV Higgs boson mass, 
 extra matters are required and they may predict a possible dark matter candidate. 
Also, the $Z\gamma$ decay was found not to be affected at one-loop level, 
 which is a distinctive prediction uncommon in other physics beyond the SM.

In this letter, 
 we continue to study the collider signature of GHU, especially the fermion decays of the SM Higgs boson. 
Since such kind of decays are proportional to yukawa coupling, 
 the dominant decay modes are the bottom decay $H \to b\bar{b}$ in the quark sector 
 and the tau decay $H \to \tau\bar{\tau}$ in the lepton sector. 
We numerically calculate the signal strength of 
 the bottom and tau decays of Higgs boson produced via the gluon fusion at the LHC in the context of GHU.

\section{The Model}
We consider an $SU(3) \times U(1)'$ GHU model in a five-dimensional flat space-time 
 compactified on $S^1/Z_2$ with the radius $R$ of $S^1$. 
The up-type quarks except for the top quark, 
 the down-type quarks and the charged leptons are embedded into ${\bf 3}$ and $\overline{{\bf 6}}$ 
 representations of $SU(3)$, respectively \cite{SSS}. 
In order to realize the large top Yukawa coupling, 
 the top quark is embedded into $\overline{{\bf 15}}$ representation of $SU(3)$ \cite{CCP}. 
The extra $U(1)'$ symmetry is required to reproduce the correct Weinberg angle, 
 and the SM $U(1)_Y$ gauge boson is given by a linear combination 
 between the gauge bosons of the $U(1)'$ and the $U(1)$ subgroup in $SU(3)$ \cite{SSS}. 
Appropriate $U(1)'$ charges for bulk fermions are assigned  
 to give the correct hypercharges for the SM fermions.

The boundary conditions are assigned 
 to reproduce the SM fields as the zero modes. 
A periodic boundary condition with respect to $S^1$ 
 is taken for all of the bulk SM fields, 
 and the $Z_2$ parity is assigned for the gauge fields and fermions 
 in the representation ${\cal R}$ 
 by using the parity matrix $P={\rm diag}(-,-,+)$ in the following. 
\ba
A_\mu (-y) = P^\dag A_\mu(y) P, \quad A_y(-y) =- P^\dag A_y(y) P,  \quad 
\psi(-y) = {\cal R}(P) \gamma^5 \psi(y) 
\label{parity}
\ea 
 where the subscripts $\mu$ ($y$) denotes the four (the fifth) dimensional component. 
With this choice of parities, 
 the $SU(3)$ gauge symmetry is broken to $SU(2) \times U(1)$. 
A $U(1)_X$ symmetry orthogonal to the hypercharge $U(1)_Y$ is anomalous in general 
 and broken at the cutoff scale, 
 which means that the $U(1)_X$ gauge boson has a mass of the cutoff scale \cite{SSS}. 
Thus, zero-mode vector bosons in the model are only the SM gauge fields.

Off-diagonal blocks in $A_y$ have zero modes as can be seen in Eq.~(\ref{parity}), 
 which corresponds to an $SU(2)$ doublet. 
In fact,  the SM Higgs doublet $H$ is identified as 
\ba
A_y^{(0)} = \frac{1}{\sqrt{2}}
\left(
\begin{array}{cc}
0 & H \\
H^\dag & 0 \\
\end{array}
\right). 
\ea
The non-zero KK modes of $A_y$ are eaten by non-zero KK modes of the SM gauge bosons 
 as their longitudinal degrees of freedom like the usual Higgs mechanism.

This parity assignment also leaves exotic massless fermions which is not included in the SM. 
Such exotic fermions are made massive 
 by introducing brane localized fermions with conjugate $SU(2) \times U(1)$ charges 
 and an opposite chirality to the exotic fermions, 
 allowing us to write brane-localized Dirac mass terms. 
These brane localized mass terms are also very important 
 to generate the flavor mixing in the context of GHU \cite{flavor GHU}.

In the GHU scenario, 
 the Yukawa interaction is given by the gauge interaction, 
 so that the mass of the SM fermions is the order of the $W$-boson mass 
 after the electroweak symmetry breaking.  
To realize light SM fermion masses, 
 one may introduce $Z_2$-parity odd bulk mass terms 
 for the SM fermions except for the top quark. 
Then, their Yukawa coupling receives exponential suppression factor 
 controlled by the bulk mass parameters $M$ such as $\exp[-\pi MR]$. 
As for the top quark Yukawa coupling, 
 the top quark should be embedded into 
 a 4-rank representation $\overline{{\bf 15}}$, for instance \cite{CCP}. 
This leads to the top quark mass as the twice of the $W$-boson mass 
 $m_t = 2 m_W$ at the compactification scale
 by a group theoretical enhancement of factor $2$ \cite{SSS}.

\section{Calculation of the signal strength of $gg \to H \to b\bar{b}, \tau\bar{\tau}$}

Our main purpose in this letter is to calculate the signal strength of $gg \to H \to b\bar{b}, \tau\bar{\tau}$. 
The Higgs production is dominated by the gluon fusion process at the LHC 
 and calculated from the coefficient of the following dimension five operator between the Higgs and the digluon, 
\ba
{\cal L}_{{\rm eff}} = C_{g} H G^a_{\mu\nu} G^{a\mu\nu} 
\ea
where $G_{\mu \nu}^a~(a=1-8)$ is the gluon field strength. 
 
The SM contribution to the Higgs production via the gluon fusion is dominated by the top quark loop as   
\begin{align}
	C_g^\SM=-\frac{1}{16\pi}\frac{\alpha_s}{v} 
	F_{1/2}\left( \left(\frac{4M_W}{m_h} \right)^2 \right)
\end{align}	
where $\alpha_s$ is the QCD fine structure constant and $m_h$ is the Higgs boson mass. 
The loop function is well known as
\begin{equation}
	F_{1/2}(\tau)=-2\tau \left[ 1+(1-\tau)(\arcsin(1/\sqrt{\tau}))^2 \right].
	\label{}
\end{equation}
In GHU, in addition to the top loop contribution, we have to take into account the KK top loop contributions, 
 which is found to be\footnote{In \cite{MO}, the terms except for the second term are missed to be considered. 
 In this letter, this point is corrected.} 
\begin{align}
	C_g^{\rm KK}=F(m_1) \times \frac{1}{2} \times 2 + F(m_2) \times 1 \times 2 + F(m_3) 
	\times \frac{3}{2} \times 1 + F(m_4) \times \frac{4}{2} \times 1
	\label{KKtop}
\end{align}
where the first factor behind $F(m_a)$ denotes the ratio for the top yukawa coupling and 
the second factor is a multiplicity of the same KK mass spectrum.    
\begin{align}
F(m_a) \equiv 
	-\frac{1}{16\pi}\frac{m_t}{v}\alpha_\s \sum_{n=1}^\infty
	\Bigg[&
		\frac{1}{m_{a+}^{(n)}}F_{1/2}\left( \left(\frac{2m_{a+}^{(n)}}{m_h} \right)^2 \right)
		- (+ \to -)
	\Bigg],
	\label{}
\end{align}
$m_t=2M_W$ and $(m_{a \pm}^{(n)})^2 = (n/R \pm a M_W)^2$.

The derivation of $C_g^{{\rm KK}}$ goes as follows. 
As discussed in \cite{MO}, 
 the ${\overline {\bf 15}}$-plet where the top quark is embedded can be decomposed under $SU(2) \times U(1)$ group as 
\ba
{\overline {\bf 15}} = {\overline{\bf 5}}_{-2/3} \oplus {\overline {\bf 4}}_{-1/6} \oplus {\overline {\bf 3}}_{1/3} 
\oplus {\overline {\bf 2}}_{5/6} \oplus {\overline {\bf 1}}_{4/3}.
\ea
The KK mass spectrum after the electroweak symmetry breaking for each decomposed representation are known to be \cite{MO}
\ba
&& 
{\overline {\bf 5}}: \left( \frac{n}{R} \pm 4 m_W \right)^2,~~ 
   \left( \frac{n}{R} \pm 2  m_W \right)^2,~~  
   \left( \frac{n}{R} \right)^2, 
  \nonumber \\ 
&& 
{\overline {\bf 4}}: \left( \frac{n}{R} \pm 3 m_W \right)^2,~~~ 
   \left( \frac{n}{R} \pm   m_W \right)^2, 
  \\ 
&& 
{\overline {\bf 3}}: \left( \frac{n}{R} \pm 2 m_W \right)^2,~~ 
   \left( \frac{n}{R} \right)^2, \quad 
{\overline {\bf 2}}: \left( \frac{n}{R} \pm m_W \right)^2, \quad 
{\overline{\bf 1}}: \left( \frac{n}{R} \right)^2. 
\nonumber
\ea  
Since all of the terms with mass splitting by the Higgs VEV contribute to the gluon fusion process 
 and yukawa coupling is given by $dm^{(0)}_{a\pm}/dv=\pm aM_W/v=\pm m_t/v \times a/2$, 
 the expression of $C_g^{{\rm KK}}$ in (\ref{KKtop}) follows in the non-zero KK sector. 
On the other hand, 
 the exotic zero mode fermions from $\overline{{\bf 3}}, \overline{{\bf 4}}$ and $\overline{{\bf 5}}$ are removed by the brane mass terms. 
Therefore, only the top quark contributes to $C_g^{{\rm SM}}$. 
The deviation of the Higgs production via the gluon fusion from the SM prediction 
 is obtained by $C^{{\rm KK}}_g/C^{{\rm SM}}_g$.

In a recent paper by the present authors \cite{AM}, 
 the deviation of the tau and the bottom yukawa couplings in GHU from the SM one has been obtained as 
\begin{align}
	\dis \frac{f}{f_\SM}
	=&\dis
	\frac{M^2-m_{\tau(b)}^2}{M^2-\pi R m_{\tau(b)}^2\sqrt{M^2-m_{\tau(b)}^2}\coth(\pi R\sqrt{M^2-m_{\tau(b)}^2})}
	\nt	\\
	&
	\times
	\dis
	\pi RM_W
	\frac{\sin\left( 2\pi RM_W \right) - 
		\left[\sin\left( 2\pi RM_W \right)-\sqrt2 \sin\left(2\sqrt{2}\pi RM_W  \right)\right]\sin^2\theta}
		{1-\cos(2\pi RM_W) - \left[\cos(2 \sqrt 2 \pi RM_W)-\cos\left( 2\pi RM_W \right)\right]\sin^2\theta}.
\end{align}
where the tau and the bottom masses are determined by the equation 
\begin{align}
	\label{KK_mass_condition2}
	&
	\dis
	\sinh^2\left[ \pi  R\sqrt{M^2-m_{\tau(b)}^2} \right]
	=
	\frac{M^2-m_{\tau(b)}^2}{m_{\tau(b)}^2} \nonumber 
	\\
	& 
	\hspace*{30mm}
	\times \left[ \sin^2\left(\pi RM_W \right)-\left( \sin^2\left( \pi RM_W \right)-\sin^2\left(\sqrt2 \pi RM_W \right) \right)\sin^2\theta \right]. 
\end{align}
The parameter $\theta$ is a mixing angle between two $SU(2)$ doublet zero modes existing per generation \cite{AM}. 
Numerical calculation of the deviation $f/f_\SM$ has been done 
 and the result was found to be almost the same as the SM prediction \cite{AM}.  

We perform a numerical calculation of the deviation of the Higgs production $C^{{\rm KK}}_g/C^{{\rm SM}}_g$ 
 in addition to our previous results for the deviation of yukawa coupling $f/f_\SM$.
The signal strength of the process $gg \to H \to b\bar{b}, \tau \bar{\tau}$ are given by
\ba
\mu = \left| \frac{C^{{\rm KK}}_g}{C^{{\rm SM}}_g} \right|^2 \times \left|  \frac{f}{f_\SM} \right|^2, 
\ea
and the numerical plots of our results are shown in Figure \ref{SS}. 
There are three parameters $R,M$ and  $\theta$ in our theory, 
 but one of them can be determined by the Eq. (\ref{KK_mass_condition2}), that is to say, 
 the combination $RM$ is determined to reproduce the realistic fermion mass. 
\def\scale{0.45}
\begin{figure}[h]
	\centering
	\includegraphics[scale=\scale]{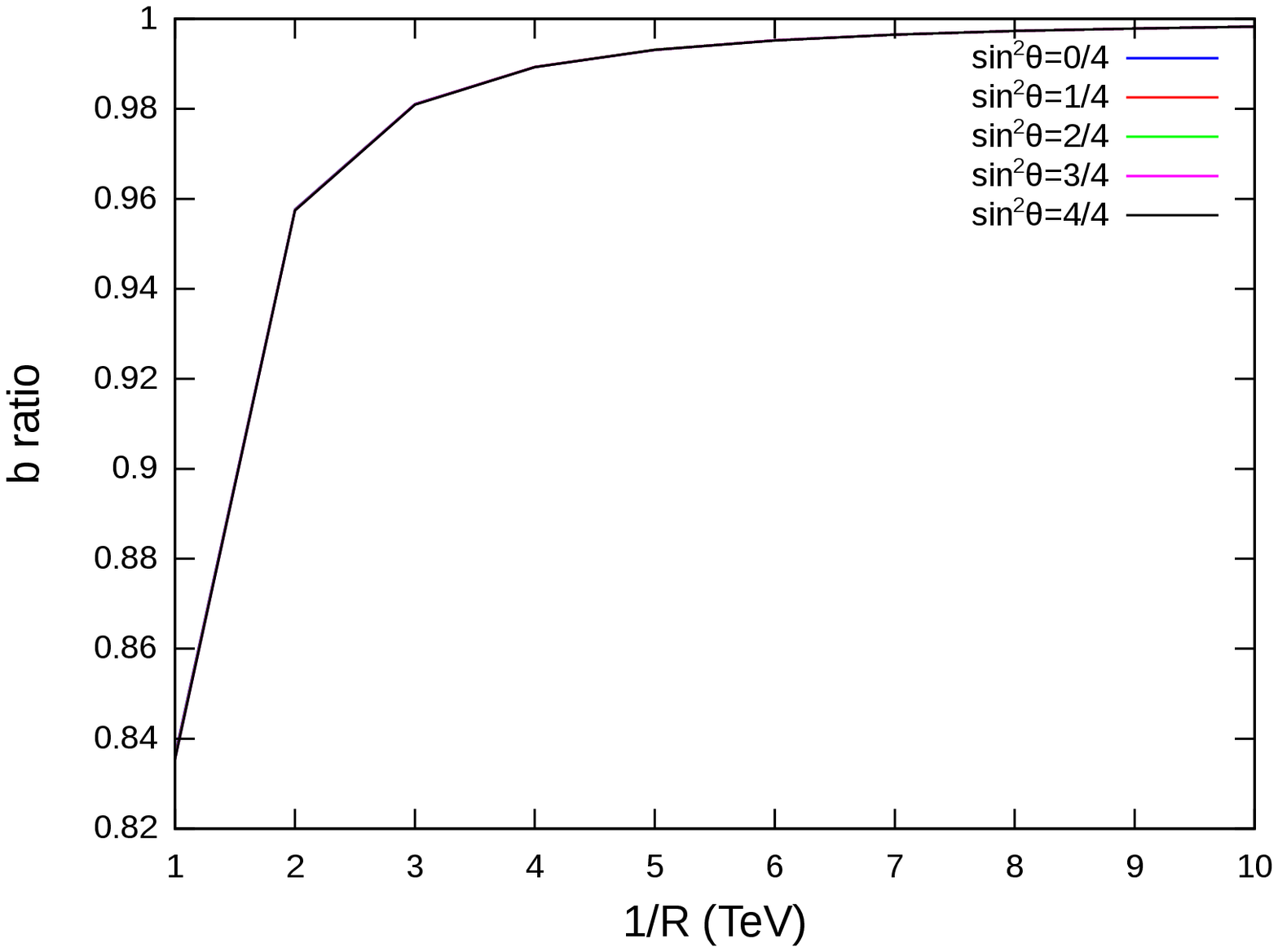}
	\includegraphics[scale=\scale]{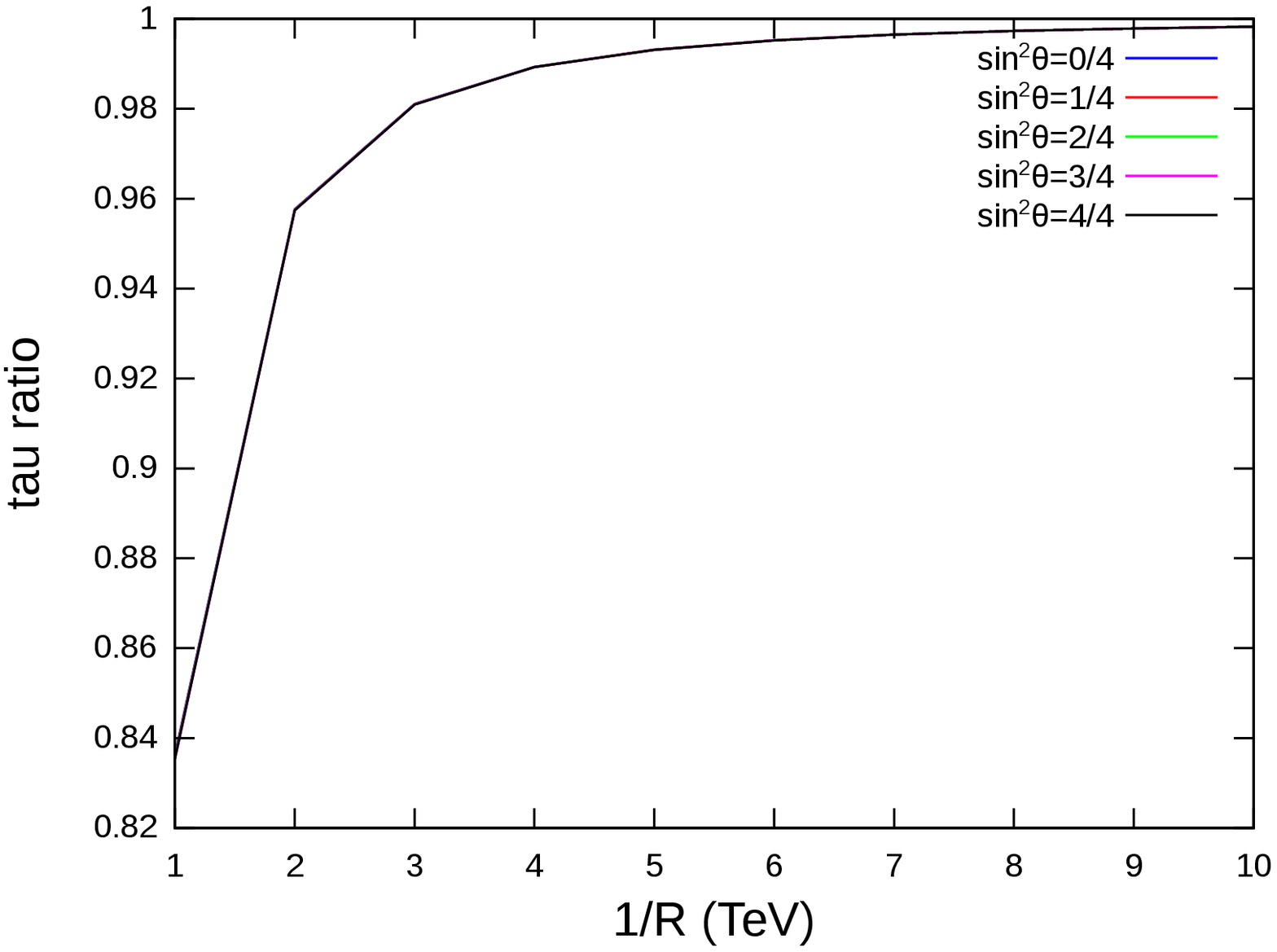}
	\caption{The left (right) plot is the signal strength of $gg \to H \to b\bar{b}(\tau \bar{\tau})$. 
	The horizontal line denotes the compactification scale. The results does not almost depend on the parameter $\theta$.}
	\label{SS}
\end{figure}
Our prediction is that the signal strength is always smaller than the unity, 
 namely the process $gg \to H \to b\bar{b}, \tau\bar{\tau}$ in GHU is always suppressed comparing to the SM prediction. 
This is because the suppression is dominantly due to the suppression of the Higgs production via the gluon fusion, 
 while the deviation of yukawa coupling is known to be very small \cite{AM}.   
This suppression nature can be distinguished from the predictions of the universal extra dimension (UED) model. 
In UED models, yukawa coupling is the same as the SM one by construction 
 and the Higgs production via the gluon fusion is enhanced by the KK mode contributions \cite{Petriello}, 
 which implies that the signal strength is larger than one. 
If the signal strength is experimentally found to be almost unity, 
 the compactificatoin scale should be around 10 TeV. 
This order of the compacttification scale is not unnatural in GHU 
 since the severe constraints from the flavor changing neutral current processes are avoided \cite{flavorGHU}. 

\section{Summary}
In this letter, we have studied the signal strength of the bottom and tau decays of the Higgs boson 
 produced via the gluon fusion at the LHC $gg \to H \to b\bar{b}, \tau\bar{\tau}$
  by taking a five dimensional $SU(3) \times U(1)'$ GHU model on the orbifold $S^1/Z_2$. 
Our generic prediction is that the signal strength is always smaller than the unity, 
 namely the process $gg \to H \to b\bar{b}, \tau\bar{\tau}$ in GHU is always suppressed comparing to the SM prediction. 
This is because the suppression is dominantly due to the suppression of the Higgs production via the gluon fusion, 
 while the deviation of yukawa coupling is known to be very small \cite{AM}.   
Our prediction can be distinguished from the UED prediction, 
 in which the signal strength is larger than unity 
 since the Higgs production via the gluon fusion is enhanced \cite{Petriello} and yukawa coupling is the same as the SM one.  
If the signal strength is experimentally found to be almost unity, 
 the compactificatoin scale should be around 10 TeV. 
This order of the compacttification scale is not unnatural in GHU 
 since the severe constraints from the flavor changing neutral current processes are avoided \cite{flavorGHU}.

We hope that our results will provide a useful information on new physics search at the LHC.

\subsection*{Acknowledgments}
The work of N.M. is supported in part by the Grant-in-Aid 
 for Scientific Research from the Ministry of Education, 
 Science and Culture, Japan No. 24540283.


\end{document}